\documentclass[aps,pra,superscriptaddress]{revtex4}

\usepackage{graphicx}
\usepackage{amsmath}
\usepackage{amssymb}
\usepackage[utf8]{inputenc}
\usepackage[english]{babel}

\usepackage{natbib}
\usepackage{color}
\usepackage{relsize}

\usepackage{bbm} 

\newcommand{\bra}[1]{\langle\,{#1}\, |}
\newcommand{\ket}[1]{|\,{#1}\,\rangle}
\newcommand{\braket}[2]{\mbox{$\langle\,{#1}\, | \,{#2}\,\rangle$}}

\newlength{\mylenunit}
\usepackage{afterpage}

\begin{document}

\title{A probability current analysis of energy transport in open quantum systems}

\author{Jan J.\ J.\ Roden}
\email{jan.roden@gmx.com}
\affiliation{Department of Chemistry, University of California, Berkeley, CA 94720, USA}

\author{K.\ Birgitta Whaley}
\affiliation{Department of Chemistry, University of California, Berkeley, CA 94720, USA}

\date{\today}

\begin{abstract}
We introduce a probability current analysis of excitation energy transfer between states of an open quantum system.
Expressing the energy transfer through currents of excitation probability between the states in a site representation enables us to gain key insights into the energy transfer dynamics.
It allows to, i)~identify the pathways of energy transport in large networks of sites and to quantify their relative weights, ii)~quantify the respective contributions of unitary dynamics, dephasing, and relaxation/dissipation processes to the energy transfer, and iii)~quantify the contribution of coherence to the energy transfer.  
Our analysis is general and can be applied to a broad range of open quantum system descriptions (with coupling to non-Markovian environments) in a straightforward manner.
\end{abstract}


\maketitle


\section{Introduction}

In the description of electronic excitation energy transport in systems of coupled sites -- such as molecular aggregates, biological and artificial light-harvesting systems, coupled quantum dots -- knowledge about the contributions of the different processes involved is important both for understanding and modeling of the energy transport mechanisms in nature, as well as for design of artificial and bio-mimetic systems that enable efficient transport.  
The key processes involved in the energy transfer that are essential for this analysis are unitary dynamics, dephasing, and relaxation/dissipation.
It has been found that the right balance between these processes can enable highly efficient, directed energy transport~\cite{rebentrost2009_033003, chin2010_065002, caruso2009_105106, roden2015psii}.
Another key aspect in the investigation of design-function relationships in natural and artificial light-harvesting systems is the identification of the specific pathways of energy transport in large networks of sites and the quantification of their relative weights~\cite{brixner2005two, ishizaki2009_17255, moix2011efficient, shibata2013_6903, bennett2013_9164, roden2015psii}.   
In such systems and for open quantum systems in general, it has long been appreciated that coherent dynamics play a role in the overall energy transfer, and the coherent features of transport have been intensively studied in a broad range of open quantum systems~\cite{haken1972coupled, haken1973exactly, silbey1976electronic, kenkre1982master, may2008charge, eisfeld2011phase, hoyer2012spatial, moix2013coherent, eisfeld2012classical}.
In recent years, experimental evidence for long-lasting coherence in biological photosynthetic systems obtained from ultra-fast spectroscopies has also sparked questions and discussions about the contribution of coherence to the energy transport that drives photosynthesis~\cite{engel2007evidence, romero2014_676, kassal2013_362, jesenko2013_174103, chenu2015coherence, olaya2008efficiency, hoyer2012spatial, kreisbeck2012long, chin2013vibrations, eisfeld2012classical, briggs2011equivalence, scholes2012_9374}.

To address these aspects of energy transfer in natural and artificial systems, we introduce here an analysis in which the energy transfer between the states of an open quantum system is expressed through excitation probability currents.
In this paper we undertake our analysis in a site basis that consists of states in which the excitation is localized on single sites, since we are interested in applying the probability current analysis to spatial energy transport. 
However, the excitation probability current analysis is general and is applicable to description of the energy transport in any basis.

Describing excitation or charge transport by means of probability currents is commonly encountered in a number of situations in condensed matter~\cite{eSilva1992prob_current, dunlap1989anderson, izzo1991calculation} and in electron transfer reactions~\cite{stuchebrukhov1996tunneling}. 
However, such a description is usually not applied in the study of light-harvesting systems, where energy transfer has instead usually been evaluated in terms of time-dependent populations of excited states (see e.g.\ Refs.~\cite{ishizaki2009_17255, Jang2014generalized, ishizaki2011interpretation, novoderezhkin2011intra, mccutcheon2011coherent, kolli2012vibrations, hoyer2012spatial, jesenko2013_174103, ritschel2011absence}).
In our analysis, we obtain the probability currents from a continuity equation for the excitation probability of the sites.
Since the total excitation probability inside the (open) system is usually conserved, such a treatment based on continuity equations and probability currents is widely applicable.
In this paper we show that this analysis enables us to address the following tasks in a straightforward manner:

\begin{enumerate}

\item Calculate the excitation probability currents between the individual sites, i.e., their direction and magnitude, thus revealing the pathways of transport.
This can be performed for a large, complicated network of sites, where the excitation of the sites couples to either a Markovian or a non-Markovian environment.

\item Quantify the respective contributions of unitary dynamics, dephasing, and relaxation/dissipation to the currents. 

\item Quantify the contribution of coherence (versus the contribution of populations) to the currents, i.e., to the energy transfer.

\end{enumerate}

Here we shall understand coherence to be quantified by the off-diagonal elements of the system's density matrix in the basis under consideration. 
As noted above, in this work we employ the site basis in order to study the spatial transport of energy.

Our treatment is applicable to the tight-binding and Holstein-like models -- whether described as open or closed quantum systems -- that are typically used for modeling excitation transfer systems of the form of biological or artificial light-harvesting systems~\cite{holstein1959model, barivsic2004holstein, romero1998holstein, cheng2009dynamics, herrera2011holstein, singh2009fluorescence}.
In the probability current framework, we obtain formulas for the contribution of the different processes -- coherence, dephasing, relaxation -- to the currents (i.e., to the excitation energy transport) from the evolution equation for the system's density matrix.
Inserting the elements of the time-dependent system density matrix into these formulas, where the matrix elements are obtained either from a numerical simulation or a tomographic reconstruction from experimental measurements, one can then numerically evaluate the contributions to the currents for specific model situations.
The analysis developed in this paper is applied in a companion paper~\cite{roden2015psii} to numerical simulations investigating excitation energy transfer and design-function relationships in the Photosystem~II protein complex that drives photosynthesis in higher plants. 

In the following, we first briefly review the description of energy transport in an open quantum system by means of a non-Markovian quantum master equation (Section~\ref{sec_qme}) to provide the underlying basis for our probability current analysis.
Then, in the main part of the paper in Section~\ref{sec_prob_current_analysis}, we introduce our probability current analysis, where we express the energy transport through probability currents and derive the different contributions of unitary evolution, dephasing, and relaxation to the currents based on the quantum master equation.
Using this framework, we then explicitly analyze the contribution of coherence to the energy transport.
To illustrate the interplay between the probability current and the coherence, we analyze the dynamical equations for the case of a Markovian environment, for which the equations are particularly simple and yield useful general insights.
A quantitative analysis for a (33 site) system with a more general non-Markovian environment is given in Ref.~\cite{roden2015psii}.
Concluding remarks are given in Section~\ref{sec_conclusion}.

\section{Quantum master equation description of energy transfer}
\label{sec_qme}

Typically, in light-harvesting systems and other systems of coupled sites featuring electronic excitation energy transfer, the electronic excitation also couples to other degrees of freedom.
For example, in molecular systems such as pigment-protein complexes in biological light-harvesting apparatuses, electronic excitation of the molecules (pigments) couples to internal vibrational modes of the molecules as well as to vibrational modes of the protein scaffold in which the pigments are embedded~\cite{damjanovic2002excitons, novoderezhkin2010physical}.  
This coupling to vibrations is often crucial in the design-function relationship, since it induces dephasing and vibrational relaxation which help to efficiently direct the energy transport to a target location (in biological systems, this might be a reaction center) where the energy is trapped and transformed~\cite{kolli2012vibrations, rebentrost2009_033003, chin2012coherence, roden2015psii}.   
It is therefore important to include such vibrational degrees of freedom in the modeling and simulation of the excitation transport, in order to take these important dephasing and relaxation effects into account. 
However, this can easily result in a large number of interacting degrees of freedom, making it challenging if not impossible to numerically treat all of these degrees of freedom explicitly.
It is thus very common to use an open quantum system description in which the problem is divided into three components.
These are, i) the ``system'' that contains the degrees of freedom of interest, i.e., for electronic excitation energy transport usually the excited electronic states of the sites, ii) the ``environment'' -- usually all the vibrational degrees of freedom -- and iii) the interaction between system and environment, i.e.\ usually the electron-vibration coupling~\cite{petruccione2002open, cheng2009dynamics}.
In this open quantum system approach one then solves an effective evolution equation in the small space of the system degrees of freedom, where the system degrees of freedom are treated explicitly and the environment degrees of freedom are taken into account only implicitly.
This approach makes even large networks of sites numerically manageable~\cite{kreisbeck2014_4045, ritschel2011absence}. 
One widely used class of effective evolution equations derived from an open quantum system approach is represented by quantum master equations.
These describe the time evolution of a reduced density matrix of the system, which formally corresponds to tracing out the environment degrees of freedom in the density matrix of the total problem~\cite{petruccione2002open, jesenko2013_174103, strunz2004convolutionless}.  
Thus, for the simulation of electronic energy transfer, the reduced density matrix usually contains the electronic degrees of freedom, i.e., where formally the partial trace over the vibrational degrees of freedom was carried out~\cite{cheng2009dynamics, ishizaki2009_17255, ritschel2011efficient}, or alternatively, the electronic and selected vibrational degrees of freedom~\cite{roden2012accounting}. 

Since such quantum master equation approaches are widely used for the description of energy transfer and are particularly popular for the modeling of light-harvesting systems~\cite{ishizaki2009_17255, kreisbeck2014_4045, caruso2009_105106, rebentrost2009_9942, ritschel2011efficient}, in this paper we shall use a quantum master equation description as the basis for our probability current analysis of the energy transfer.
In this section, we therefore briefly outline the quantum master equation description, before introducing the probability current analysis in Section~\ref{sec_prob_current_analysis}. 

We consider electronic excitation energy transport in a system of coupled sites, and as a dynamical model consider a quantum master equation of the form 
\begin{equation}
  \label{qme}
  \partial_t\rho(t) = \mathcal{P}(\rho(t)) = -i[H, \rho(t)] + \mathcal{L}_{\rm non-unitary}(\rho(t)),
\end{equation}
that describes the time evolution of the reduced density matrix $\rho(t)$ of the system, which contains the electronic degrees of freedom of the sites, with initial condition $\rho_0$, with $H$ the system Hamiltonian that contains the energies of the sites and the couplings between these.
(Here and throughout the paper we set $\hbar\equiv 1$).
The first term in Equation~(\ref{qme}) describes unitary dynamics of the system and the second term describes the non-unitary contribution to the system dynamics due to interaction of the system with the environment.
The probability to find electronic excitation localized on site $n$ at time $t$ is then given by 
\begin{equation}
  \rho_{nn}(t) = \bra{n}\rho(t)\ket{n},
\end{equation}
where $\ket{n}$ is the state in which only site $n$ is excited and all other sites are in the ground state.
Thus, the states $\ket{n}$ span the one-excitation subspace.
The transport of electronic excitation energy between the sites will manifest itself in the time-dependent behavior of these excitation probabilities $\rho_{nn}(t)$ of the sites.

We assume that the non-unitary term of Equation~(\ref{qme}) is a sum of two contributions, one of dephasing and one of dissipation/relaxation between electronic states due to the coupling to the environment:
\begin{equation}
  \label{dephas_and_relax}
  \mathcal{L}_{\rm non-unitary}(\rho(t)) = \mathcal{L}^{\rm Dephas}(\rho(t)) + \mathcal{L}^{\rm Relax}(\rho(t)).
\end{equation}
We next consider the form of these dephasing and electronic relaxation contributions.
For our dynamical model, we shall assume that Equation~(\ref{qme}) is realized by a convolutionless, non-Markovian quantum master equation, developed specifically to treat non-Markovian interactions with the environment~\cite{strunz2004convolutionless}. 
Such an equation has been used to describe energy transfer in molecular systems where the electronic excitation couples to non-Markovian vibrational modes of the environment~\cite{ritschel2011efficient, ritschel2011absence, roden2015psii}.
Then the dephasing contribution can be written in the form
\begin{equation}
  \label{non_markov_dephas}
  \mathcal{L}^{\rm Dephas}(\rho(t)) = \sum_n \left(L_n^{\rm D}\rho(t) {A_n^{\rm D}}^{\dagger}(t) + A_n^{\rm D}(t)\rho(t) {L_n^{\rm D}}^{\dagger} - {L_n^{\rm D}}^{\dagger}A_n^{\rm D}(t)\rho(t) - \rho(t) {A_n^{\rm D}}^{\dagger}(t)L_n^{\rm D} \right),
\end{equation}
where the non-Markovian influence of the environment is captured by time-dependent auxiliary operators $A_n^{\rm D}(t)$, which follow a separate evolution equation that is independent of the density matrix $\rho(t)$ (see Refs.~\cite{strunz2004convolutionless, ritschel2011efficient, roden2015psii}). 
The system operators 
\begin{equation}
  \label{dephas_op}
  L_n^{\rm D} = \ket{n}\bra{n}
\end{equation}
couple electronic excitation of site $n$ to the environment.
We note that the coupling to the non-Markovian environment that is described by Eqs.~(\ref{non_markov_dephas}) and~(\ref{dephas_op}) not only induces dephasing, but can also induce {\it relaxation} within the vibrational manifold of the environment~\cite{roden2012accounting}.
However, in the remainder of the paper we will refer to the term Eq.~(\ref{non_markov_dephas}) simply as ``dephasing'', to distinguish it from the {\it electronic} relaxation (of the electronic states) that we describe next.

Electronic dissipation/relaxation between electronic states is analogously described through
\begin{equation}
  \label{non_markov_relax}
  \mathcal{L}^{\rm Relax}(\rho(t)) = \sum_{l,n\neq l} \left(L_{nl}^{\rm R}\rho(t) {A_{nl}^{\rm R}}^{\dagger}(t) + A_{nl}^{\rm R}(t)\rho(t) {L_{nl}^{\rm R}}^{\dagger} - {L_{nl}^{\rm R}}^{\dagger}A_{nl}^{\rm R}(t)\rho(t) - \rho(t) {A_{nl}^{\rm R}}^{\dagger}(t)L_{nl}^{\rm R} \right),
\end{equation}
where the system coupling operators
\begin{equation}
  \label{relax_op}
  L_{nl}^{\rm R} = \ket{l}\bra{n}
\end{equation}
describe relaxation from a state $\ket{n}$ to a state $\ket{l}$, and the corresponding time-dependent auxiliary operators $A_{nl}^{\rm R}(t)$ similarly include the non-Markovian effects of the coupling to the environment. 

In the Markovian limit, where the correlation time of the environment is assumed to be short compared to the relevant system time scales of the dynamics, the time-dependent auxiliary operators for dephasing become time-independent~\cite{strunz2004convolutionless}: 
\begin{equation}
  \label{markov_limit_dephas}
  A_n^{\rm D}(t) \rightarrow \frac{1}{2}\gamma_n^{\rm D} L_n^{\rm D}
\end{equation}
where $\gamma_n^{\rm D}$ are the system-environment coupling parameters.
For the relaxation contribution we have analogously~\cite{strunz2004convolutionless}
\begin{equation}
  \label{markov_limit_relax}
  A_{nl}^{\rm R}(t) \rightarrow \frac{1}{2}\gamma_{nl}^{\rm R} L_{nl}^{\rm R}.
\end{equation}
In this limit, the non-Markovian quantum master equation~(\ref{qme}) becomes the well-known Markovian Lindblad equation~\cite{strunz2004convolutionless}.
This can be easily seen by inserting Eqs.~(\ref{markov_limit_dephas}) and~(\ref{markov_limit_relax}) in the terms for dephasing and relaxation Eqs.~(\ref{non_markov_dephas}) and~(\ref{non_markov_relax}), respectively.

\section{Probability current analysis of energy transfer}
\label{sec_prob_current_analysis}

Our goal is to analyze the energy transfer in terms of probability currents between the sites, in order to i)~provide information about the transfer pathways, ii)~reveal the respective contributions of unitary dynamics, dephasing, and relaxation to the transport, and iii)~identify the contribution of coherence between the sites to the transport.
In the following, we will develop this analysis based on the non-Markovian quantum master equation description of the previous section.

Energy transfer between the sites occurs when the electronic excitation is transferred between the sites, driven by the inter-site coupling.
Since the overall excitation probability in the system is conserved, $\sum_{n}\rho_{nn}(t) = 1$, even if the system is open (i.e., it couples to an environment and $\rho(t)$ is identified with the reduced density matrix of the system), and a continuity equation holds,
\begin{equation}
\label{conti_eq}
  \partial_t\rho_{nn}(t) = \sum_{l\neq n} j_{ln}(t),
\end{equation}
where $j_{ln}(t)$ is the net probability current at time $t$ that transports excitation probability from a site $l$ to site $n$.
If $j_{ln}(t)$ is positive, excitation is transported from site $l$ to site $n$; 
if $j_{ln}(t)$ is negative, there is transport from site $n$ to site $l$.
These currents can be identified with the energy transfer between the sites.
(We note that instead of the probability currents considered in the present work, others have considered {\it energy} currents to quantify excitation energy transfer~\cite{nalbach2011exciton}. 
However, this can be problematic in open systems, since the energy inside an open system is {\it not} a conserved quantity, and therefore there is no corresponding continuity equation.)

By making use of the additivity of currents $j_{ln}(t)$ between the sites, the currents between sub-complexes that consist of a number of sites can also be calculated.
Thus, the current $J_{AB}(t)$ between a sub-complex $A$ and a sub-complex $B$ is given by
\begin{equation}
  J_{AB}(t) = \sum_{l\in A}\sum_{n\in B} j_{ln}(t). 
\end{equation}
When the current $J_{AB}(t)$ is positive, there is a net flow from $A$ to $B$, and when $J_{AB}(t)$ is negative, there is a net flow from $B$ to $A$.

We now analyze the energy transfer dynamics given by the quantum master equation~(\ref{qme})--(\ref{relax_op}) of Section~\ref{sec_qme} in terms of these probability currents.
We find that this quantum master equation leads to changes of the site populations in time that are given by
\begin{alignat}{3}
\label{pop_change_qme}
  \partial_t\rho_{nn}(t) & =  \ -i\bra{n}[H, \rho(t)]\ket{n} & {} + {} & \ \ \bra{n}\mathcal{L}^{\rm Dephas}(\rho(t))\ket{n}\ & {} + {} & \ \ \bra{n}\mathcal{L}^{\rm Relax}(\rho(t))\ket{n}\nonumber \\[5pt]
                       & =  \ \sum_{l\neq n} 2 H_{ln}\,{\rm Im}(\rho_{ln}(t))\  & {} + {} & \ \ \sum_{l\neq n}\, 0                   & {} + {} & \ \ \sum_{l\neq n} 2\,{\rm Re}\left\{\bra{l}\left(\rho(t){A_{ln}^{\rm R}}^{\dagger}(t) - A_{nl}^{\rm R}(t)\rho(t)\right)\ket{n}\right\},
\end{alignat}
where the elements $H_{ln}=\bra{l}H\ket{n}$ of the Hamiltonian are assumed to be real.
Comparing Equation~(\ref{pop_change_qme}) with the continuity equation~(\ref{conti_eq}), we find that the population currents from sites $l$ to sites $n$ are given by
\begin{equation}
\label{total_current}
  j_{ln}(t) \  = \ j_{ln}^{\rm Unitary}(t) \ + \ j_{ln}^{\rm Dephas}(t) \ + \ j_{ln}^{\rm Relax}(t),
\end{equation}
with
\begin{equation}
  \label{contrib_total_current}
  \begin{split}
    & j_{ln}^{\rm Unitary}(t)  =  2 H_{ln}\,{\rm Im}(\rho_{ln}(t)),\\
    & j_{ln}^{\rm Dephas}(t)\,  =  0,\\
    & j_{ln}^{\rm Relax}(t)\ \  =  2\,{\rm Re}\left(\sum_k\rho_{lk}(t){A_{ln, nk}^{\rm R}}^{*}(t) - A_{nl, lk}^{\rm R}(t)\rho_{kn}(t)\right),
  \end{split}
\end{equation}
for $l \neq n$.
The currents are a sum of the three contributions of unitary dynamics, dephasing, and electronic relaxation.
In Equation~(\ref{contrib_total_current}), these contributions are expressed by the coherences $\rho_{ln}(t)$ between the sites and the populations $\rho_{nn}(t)$ of the sites. 
$H_{ln}$ are the inter-site couplings and $A_{nl, lk}^{\rm R}(t) = \bra{l}A_{nl}^{\rm R}(t)\ket{k}$ are the site-basis matrix elements of the auxiliary operators describing relaxation between states $n$ and $l$ (${*}$ denotes complex conjugation).
Given a system density matrix $\rho(t)$ and the relaxation auxiliary operators $A_{nl}^{\rm R}(t)$ at a time $t$, one can then use Equation~(\ref{contrib_total_current}) to calculate the currents between the sites.
We now discuss the three different contributions to the total current of Eqs.~(\ref{total_current}), (\ref{contrib_total_current}).

\bigskip
\noindent{\bf Unitary contribution:}\smallskip

As can be seen in Equation~(\ref{contrib_total_current}), the unitary contribution $j_{ln}^{\rm Unitary}(t)$ to the total current is caused {\it entirely} by the coherence between the sites, in particular by the imaginary components of this, since there are no diagonal elements of $\rho(t)$ (site populations) present in this term.
This is an important result, since it shows that without coherence the unitary contribution to the energy transfer would be zero.
As expected, this contribution is proportional to the inter-site couplings $H_{ln}$.
On the other hand, this unitary contribution of the currents does not explicitly depend on the diagonal elements of $H$, i.e., the energies of the sites.
Nevertheless, as we will see in the following, $j_{ln}^{\rm Unitary}(t)$ does {\it implicitly} depend on the site energies through the time evolution of the coherence.
The above result for unitary contribution $j_{ln}^{\rm Unitary}(t)$ is well known, and is given for a system with nearest-neighbor interaction in, e.g.\ Ref.~\cite{eSilva1992prob_current}.

An important property of this unitary contribution $j_{ln}^{\rm Unitary}(t)$ is that, because the coherence $\rho_{ln}(t)$ is limited by the Cauchy-Schwarz inequality
\begin{equation}
  \label{cauchy_schwarz}
  |\rho_{ln}(t)|^2 \leq \rho_{ll}(t)\,\rho_{nn}(t), 
\end{equation}
for the coherence and population of a density matrix $\rho(t)$, the unitary contribution to the current is also bounded:
\begin{equation}
  \label{limited_unitary_current}
  |j_{ln}^{\rm Unitary}(t)| = 2 |H_{ln}{\rm Im}(\rho_{ln}(t))| \leq  2 |H_{ln}| \sqrt{\rho_{ll}(t)\,\rho_{nn}(t) - \left({\rm Re}(\rho_{ln}(t))\right)^2}.
\end{equation}
Thus, the larger the real component of the coherence between two sites, the more limited is the unitary current between these sites.
It is important to emphasize the two different roles that the imaginary and the real components of the coherence play in the energy transfer:
while the imaginary component of the coherence constitutes the actual transport current, the real component does nevertheless have a constraining effect on the energy transfer.
Because of these two different roles, it can be insightful to study the imaginary and real components of the coherence involved in energy transfer dynamics separately.
In Section~\ref{subsec:Lindblad} we do this explicitly for the special case of system-environment coupling in the Markovian limit.

\bigskip
\noindent{\bf Dephasing contribution:}\smallskip

In Equation~(\ref{total_current}), the contribution of dephasing to the current is zero, regardless of the specific form of the dephasing, i.e., the specific behavior of the time-dependent auxiliary operators $A_n^{\rm D}(t)$ that describe the non-Markovian influence of the coupling to the environment. 
(Here we have assumed that $\braket{n}{l}=\delta_{nl}$, i.e., the states are orthonormal.)
This finding is important, because it means that the dephasing does not influence the currents and hence the energy transfer explicitly.
Thus, in a model that only includes unitary dynamics and non-Markovian dephasing, but no electronic relaxation between some or all of the sites -- a model often applied~\cite{ishizaki2009_17255, kreisbeck2011_2166, ritschel2011efficient} -- the currents between the respective sites would be given entirely by the unitary contribution $j_{ln}^{\rm Unitary}(t)$. 
Nevertheless, from Equation~(\ref{non_markov_dephas}) it follows that the dephasing term $\mathcal{L}^{\rm Dephas}(\rho(t))$ does act on the coherence between the sites, via the terms
\begin{equation}
  \label{elements_non_markov_dephas}
  \bra{l}\mathcal{L}^{\rm Dephas}(\rho(t))\ket{n} = \bra{l}\left(\rho(t){A_l^{\rm D}}^{\dagger}(t) - \rho(t) {A_n^{\rm D}}^{\dagger}(t) - A_l^{\rm D}(t)\rho(t) + A_n^{\rm D}(t)\rho(t)\right)\ket{n}\,(1 - \delta_{ln}).
\end{equation}
Thus, the dephasing term can, for example, cause the coherence to decay.
Therefore, the dephasing can indirectly influence the unitary contribution $j_{ln}^{\rm Unitary}(t)$ to the current that is driven by the coherence, and thus implicitly influence the energy transport.
Since the action of the dephasing terms on the coherence depends on the time-dependent auxiliary operators $A_n^{\rm D}(t)$, the precise way in which coherence is influenced will depend on the details of the non-Markovian dynamics of the environment.

In the Markovian limit however, the action of the dephasing terms on the coherence is simple.
In this limit, where $A_n^{\rm D}(t) \rightarrow \frac{1}{2}\gamma_n^{\rm D} L_n^{\rm D}$ (Eq.~(\ref{markov_limit_dephas})), the dephasing is described by Lindblad terms,
\begin{equation}
  \label{coher_decay_lindblad}
  \bra{l}\mathcal{L}_{\rm Lindbl}^{\rm Dephas}(\rho(t))\ket{n} = -\frac{1}{2}(\gamma_l^{\rm D} + \gamma_n^{\rm D})\,\rho_{ln}(t)\,(1 - \delta_{ln})
\end{equation}
that simply cause the coherences between the sites to decay on a time scale given by the coupling parameters $\gamma_n^{\rm D}$.
Therefore, the dephasing can diminish or inhibit the unitary contribution $j_{ln}^{\rm Unitary}(t)$ that is driven by the coherence.
It is well known that for dephasing that is fast compared to the other timescales of the dynamics, the energy transfer is inhibited (Quantum Zeno effect)~\cite{rebentrost2009_033003}.  

\bigskip
\noindent{\bf Relaxation contribution:}\smallskip

Equation~(\ref{contrib_total_current}) shows that the third term $j_{ln}^{\rm Relax}(t)$, the contribution of electronic relaxation to the total current, depends on both coherence and populations of the sites and also on the time-dependent auxiliary operators.
Therefore, the contribution of this term to the total current can have a complicated dependence on the interplay of electronic coherence and populations with the environment.

In the Markovian limit, on the other hand, the situation is again much simpler.
Inserting $A_{nl}^{\rm R}(t) \rightarrow \frac{1}{2}\gamma_{nl}^{\rm R} L_{nl}^{\rm R}$ (Eq.~(\ref{markov_limit_relax})) into the $j_{ln}^{\rm Relax}(t)$ term of Equation~(\ref{contrib_total_current}), yields the Markovian relaxation contribution
\begin{equation}
  \label{lindblad_relax_current}
  j_{ln}^{\rm Relax, Lindbl}(t) = \gamma_{ln}^{\rm R}\rho_{ll}(t) - \gamma_{nl}^{\rm R}\rho_{nn}(t).
\end{equation}
We see that this contribution now has the character of a purely classical rate equation (see, e.g., Ref.~\cite{zia2010towards}), where $\gamma_{ln}^{\rm R}$ specifies the rate of electronic relaxation transport from a state $\ket{l}$ to a state $\ket{n}$ and $\gamma_{nl}^{\rm R}$ is the rate for the reverse process.
This relaxation contribution to the excitation transfer relies entirely on the populations; coherence between the sites does not enter into this process. 
Since this current contribution caused by the relaxation terms does not depend on coherence between the sites, it will not be destroyed by site dephasing.
Quantum master equations with non-Markovian dephasing terms, in combination with such {\it Markovian} Lindblad terms for electronic relaxation/dissipation have been used for the description of energy transfer in light-harvesting systems, where the Lindblad terms describe (irreversible) trapping of the energy (e.g.\ in a reaction center of a biological light-harvesting apparatus)~\cite{kreisbeck2011_2166, caruso2009_105106, jesenko2013_174103, roden2015psii}.

\subsection{Quantifying the contribution of coherence to energy transfer}

The above analysis has shown that in the absence of electronic relaxation, the excitation transfer during a time interval $\Delta t$ depends entirely on the coherence between the sites, given by the off-diagonal elements of the density matrix $\rho(t)$ in this time interval.
As described by Equation~(\ref{contrib_total_current}), this constitutes the unitary contribution $j_{ln}^{\rm Unitary}(t)$ to the current.
If this coherence is zero, there will be no excitation transfer during this time interval.

We can quantify the contribution of coherence to the energy transfer, i.e., to the currents $j_{ln}(t)$, by writing the currents as a sum of the two contributions of the populations of the sites and the coherence between the sites
\begin{equation}
  j_{ln}(\rho(t)) = j_{ln}^{\rm pop}(\rho^{\rm d}(t)) + j_{ln}^{\rm coher}(\rho^{\rm nd}(t)).
\end{equation}
Here the first term contains the diagonal elements of the density matrix at time $t$ in the site basis (populations) and the second term contains the off-diagonal elements (coherences), i.e., $\rho(t) = \rho^{\rm d}(t) + \rho^{\rm nd}(t)$.
This partitioning is possible because the propagator $\mathcal{P}$ of the evolution equation~(\ref{qme}) is linear in $\rho(t)$.
The same partitioning can be applied to the changes $\partial_t\rho_{nn}(t)$ of the populations of the sites, which are just the sums of the currents to/from the sites (see Eq.~(\ref{conti_eq}))
\begin{equation}
  \partial_t\rho_{nn}(t) = \sum_{l\neq n} j_{ln}^{\rm pop}(\rho^{\rm d}(t)) + \sum_{l\neq n} j_{ln}^{\rm coher}(\rho^{\rm nd}(t)) = \bra{n}\mathcal{P}(\rho^{d}(t))\ket{n} + \bra{n}\mathcal{P}(\rho^{nd}(t))\ket{n}.
\end{equation}
Here the contributions of population and coherence can be readily calculated by (numerically) applying the propagator $\mathcal{P}$ of the evolution equation~(\ref{qme}) to the diagonal and off-diagonal part of a given density matrix $\rho(t)$ at each point in time, respectively.

In the case of non-Markovian dephasing, but Markovian electronic relaxation/dissipation, described above (see Eq.~(\ref{lindblad_relax_current})), we have the simple partitioning 
\begin{equation}
  j_{ln}^{\rm coher}(\rho^{\rm nd}(t)) = j_{ln}^{\rm Unitary}(t) \ \ \ \ \mbox{ and } \ \ \ \ j_{ln}^{\rm pop}(\rho^{\rm d}(t)) = j_{ln}^{\rm Relax, Lindbl}(t),
\end{equation}
since the unitary contribution contains only coherence and the contribution of the electronic relaxation contains only populations and the dephasing contribution is zero.
This is an important case, since it applies to many of the commonly used dynamical models for energy transfer in light-harvesting systems~\cite{kreisbeck2011_2166, jesenko2013_174103, roden2015psii}.
Again it should be emphasized that without coherence, the unitary current, which is often the main contribution in excitation energy transfer, vanishes and is therefore completely reliant on the presence of coherence.

We note that there are other dynamical models commonly applied to describe energy transfer in which coherence is not explicitly present, e.g., the (generalized) F\"{o}rster method or other methods using classical master equations (rate equations)~\cite{raszewski2008_4431, bennett2013_9164}.
This is not a contradiction to the aforementioned result that coherence is essential for the occurrence of energy transfer (in the absence of transfer via electronic relaxation between the sites), because such classical rate equations can be seen as {\it effective} models derived from a general quantum master equation.
When this reduction is done consistently, it is evident that such classical rate equations implicitly take coherence into account, even though no coherence terms appear explicitly in the classical rate equation.
A thorough analysis of this reduction is given in Refs.~\cite{jesenko2013_174103, jesenko2014comparison}. 
For example, a classical rate equation containing conventional F\"{o}rster rates for the excitation energy transfer (EET) is valid only within a restricted parameter regime where the electronic inter-site coupling is weak compared to the electron-vibration coupling, providing a limiting case of the general quantum master equation. 
Since, as demonstrated above, in the quantum master equation model inter-site coherence is necessary for EET regardless of the parameter regime, it follows that an effective F\"{o}rster model {\it implicitly} takes this coherence into account, even though no coherence terms appear explicitly in the F\"{o}rster rate equation.
We note that in the literature, F\"{o}rster transfer is often termed ``incoherent'' (see, e.g., Refs.~\cite{wilkins2014coher_fmo, kassal2013_362, chachisvilis1997excitons}) since it does not contain explicit coherence terms.
However, this is somewhat misleading, since F\"{o}rster transfer does implicitly take coherence into account.

\subsection{Energy transfer pathways} 

Aside from the analysis of the contributions of the different processes to the probability currents, the probability current description is also very useful to identify the pathways of energy transfer in a network of sites and to reveal the relative weights, i.e., importance, of the different pathways.
This can provide important insight into the design-function relationships of biological systems such as light-harvesting complexes, inter alia.
To reveal the pathways, one can integrate the probability currents between the sites over a certain time interval of interest, to obtain the direction of transport and the net amount of excitation probability that has been transported via each pathway during this time period.
Thus, the net amount of probability $\Delta P_{ln}(\Delta t)$ that has been transported between a site $l$ and a site $n$ during the time $\Delta t$ can be calculated from
\begin{equation}
  \label{pathway_pop}
  \Delta P_{ln}(\Delta t) = \int_{t_0}^{t_0 + \Delta t} dt\, j_{ln}(t),
\end{equation} 
where the currents $j_{ln}(t)$ are calculated beforehand through Equations~(\ref{total_current}), (\ref{contrib_total_current}) (or through an analogous equation based on a different underlying dynamical model), from a given time-dependent density matrix $\rho(t)$ that was itself calculated beforehand through propagation with the same dynamical model, e.g.\ the quantum master equation described in Section~\ref{sec_qme}. 

\begin{figure}
\centering
\includegraphics[width=1.0\mylenunit]{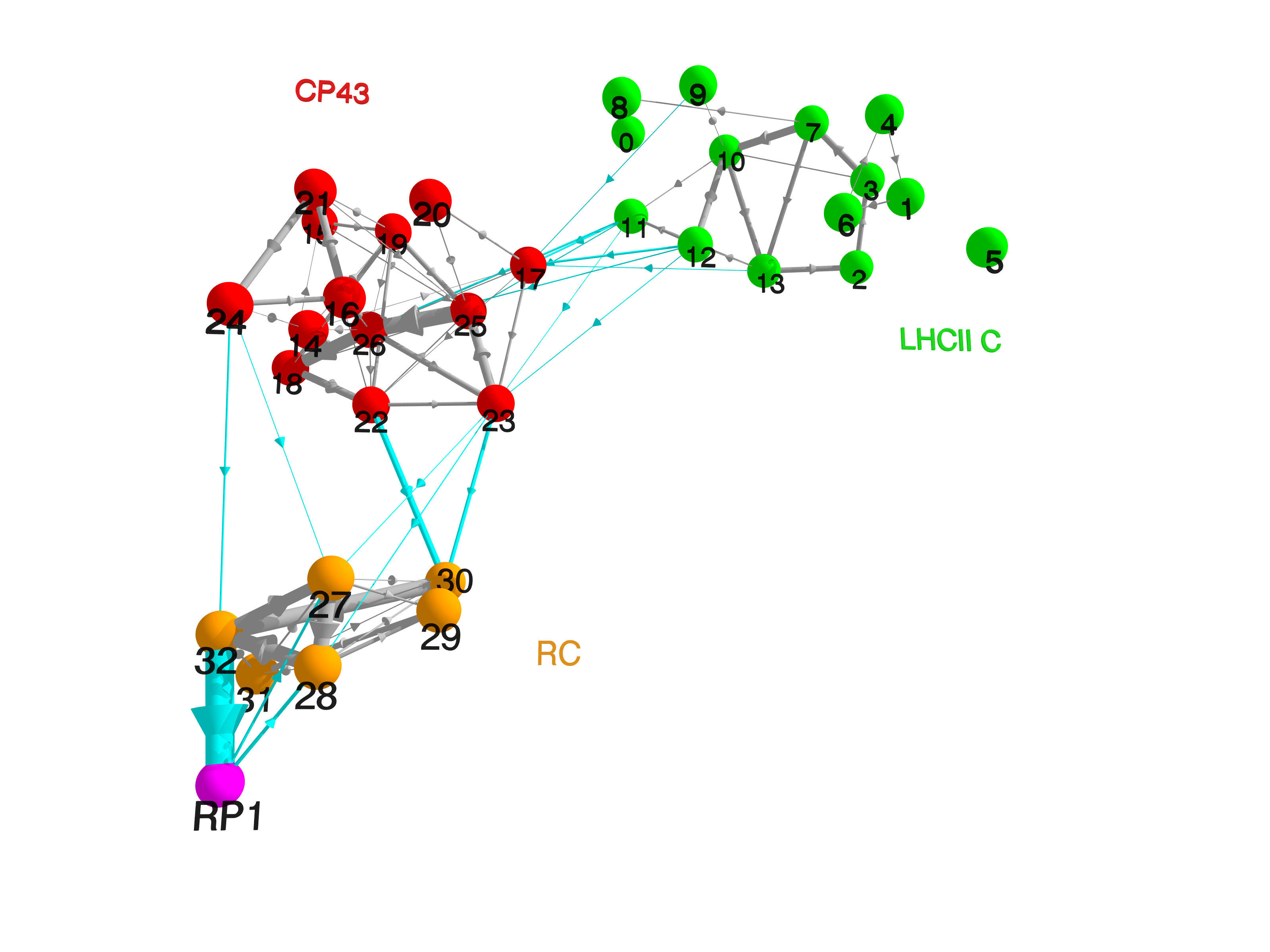}

\caption{Diagram from Ref.~\cite{roden2015psii} that shows the time-integrated excitation probability currents between the pigments in a Photosystem~II super-complex (over a time period of 1~ns), based on a numerical propagation of the electronic density matrix using a non-Markovian quantum master equation.
(Initially only pigments~7 and~10 are excited.)
The arrows show the directions and their thickness the relative magnitude of the integrated currents.
(See Ref.~\cite{roden2015psii} for details of the simulation.)
}
\label{fig_PSII_currents}
\end{figure}

This approach is used in a companion paper~\cite{roden2015psii} to reveal the pathways of energy transport in the Photosystem~II super-complex, a light-harvesting apparatus driving photosynthesis in higher plants, and to gain insight into the design-function relationships of this important photosynthetic apparatus. 
To illustrate this example, one of the diagrams of Ref.~\cite{roden2015psii} showing the transport pathways (time-integrated probability currents) between outer (antenna) and inner (reaction center) sites (pigments) of one of the primary photosynthetic systems in green plants (Photosystem~II) is shown in Figure~\ref{fig_PSII_currents}.
The arrows in the diagram show the directions and relative importance (magnitude of the integrated current) of the different pathways.
See Ref.~\cite{roden2015psii} for a full analysis.

\subsection{Dynamics for Markovian system-environment coupling}
\label{subsec:Lindblad}

When the memory time of the environment is short relative to the characteristic time scale of the system dynamics, the coupling to the environment is Markovian and the energy transfer dynamics are particularly simple.
Therefore, it is instructive to take a closer look at this specific situation.
We show below that in this Markovian setting, only three quantities are enough to describe the dynamics.
These are, i) the probability current between the sites, ii) the real component of the coherence between the sites, and iii) the imaginary component of the coherence. 

In the Markovian limit, the quantum master equation of Section~\ref{sec_qme} reduces to a Lindblad equation, with Equations~(\ref{markov_limit_dephas}) and~(\ref{markov_limit_relax}) applying to the dephasing and relaxation terms.
From Equations~(\ref{total_current}), (\ref{contrib_total_current}), and~(\ref{lindblad_relax_current}), the total population current from site $l$ to site $n$ is then given by
\begin{equation}
  \label{lindblad_total_current}
  j_{ln}^{\rm Lindbl}(t) = - \partial_t\, d_{ln}(t) = j_{ln}^{\rm Unitary}(t) + j_{ln}^{\rm Relax, Lindbl}(t) = 2 V_{ln}\,{\rm Im}(\rho_{ln}(t)) + \left(\gamma_{ln}^{\rm R}\rho_{ll}(t) - \gamma_{nl}^{\rm R}\rho_{nn}(t)\right),       
\end{equation}
where $d_{ln}(t)\equiv (\rho_{ll}(t) - \rho_{nn}(t))/2$ is the difference of the populations of sites $l$ and $n$, and $V_{ln} = H_{ln}$ is the electronic coupling between the sites.
The unitary contribution to the population current is driven by the imaginary component ${\rm Im}(\rho_{ln}(t))$ of the coherence between the sites and the relaxation contribution depends on the population $\rho_{nn}(t)$ of the sites.
Thus, we consider the evolution equations for the imaginary and real components of the coherence separately.
From the Lindblad dynamics (Eq.~(\ref{qme}) with Eqs.~(\ref{markov_limit_dephas}), (\ref{markov_limit_relax}) inserted in Eqs.~(\ref{non_markov_dephas}), (\ref{non_markov_relax})), we obtain for the imaginary component of the coherence
\begin{equation}
  \label{evol_imag_coher_lindblad}
  \partial_t\,{\rm Im}(\rho_{ln}(t)) = 2V_{ln}\,d_{ln}(t) - \Delta_{ln}\,{\rm Re}(\rho_{ln}(t)) - \Gamma_{ln}\,{\rm Im}(\rho_{ln}(t)) + \sum_{k\neq l,n}\left(V_{kn}\,{\rm Re}(\rho_{lk}(t)) - V_{lk}\,{\rm Re}(\rho_{kn}(t))\right),
\end{equation}
where $\Delta_{ln}\equiv (H_{ll} - H_{nn})$ is the energy gap between the transition energies of sites $l$ and $n$.
The dephasing and electronic relaxation rates are combined into a single decay rate $\Gamma_{ln}\equiv (\gamma_l^{\rm D} + \gamma_n^{\rm D} + \sum_{k\neq l}\gamma_{lk}^{\rm R} + \sum_{k\neq n}\gamma_{nk}^{\rm R})/2$.
For the real component of the coherence, we obtain
\begin{equation}
  \label{evol_real_coher_lindblad}
  \partial_t\,{\rm Re}(\rho_{ln}(t)) = \Delta_{ln}\,{\rm Im}(\rho_{ln}(t)) - \Gamma_{ln}\,{\rm Re}(\rho_{ln}(t)) - \sum_{k\neq l,n}\left(V_{kn}\,{\rm Im}(\rho_{lk}(t)) - V_{lk}\,{\rm Im}(\rho_{kn}(t))\right).
\end{equation}
The three coupled equations~(\ref{lindblad_total_current})--(\ref{evol_real_coher_lindblad}) completely describe the energy transfer dynamics.
Since the structure of these equations is very simple, some qualitative observations can be readily read off from these equations.

First, we note that the imaginary component of the coherence creates population transport via the first term in Equation~(\ref{lindblad_total_current}), as discussed above.
The population transport in turn couples back into the imaginary component of the coherence via the population difference $d_{ln}(t)$ in the first term in Equation~(\ref{evol_imag_coher_lindblad}).
The real component of the coherence (Equation~(\ref{evol_real_coher_lindblad})), on the other hand, is not directly coupled to the population transport.
Nevertheless, it influences the transport of population indirectly, because it is coupled to the imaginary component of the coherence via the first and last terms in Equation~(\ref{evol_real_coher_lindblad}) and the second and last terms in Equation~(\ref{evol_imag_coher_lindblad}).
These terms that couple the real and imaginary components of the coherence can be divided into two distinct contributions.
First, terms depending on the {\it local} coherence, i.e., the coherence between two sites $l$ and $n$ that affect the population current between the same two sites (the second term in Equation~(\ref{evol_imag_coher_lindblad}) and the first term in Equation~(\ref{evol_real_coher_lindblad})).
Second, terms that account for the non-local influence of the coherence stemming from the coupling to other sites (the last terms of Eqs.~(\ref{evol_imag_coher_lindblad}) and~(\ref{evol_real_coher_lindblad}), respectively). 
The first set of terms taking the local coherence into account scale with the energy gaps $\Delta_{ln}$ between the sites, while the second set of terms describing the non-local influence of coherence scale with the electronic inter-site couplings $V_{kn}$.

We note further that the influence of the energy gaps $\Delta_{ln}$ on the population current depends on the magnitude of the real component of the coherence since it appears only as a factor in the ${\rm Re}(\rho_{ln}(t))$ term in Equation~(\ref{evol_imag_coher_lindblad}).

Another important property of the Lindblad dynamics is that the Markovian dephasing and relaxation leads to a simple decay term for the imaginary component of the coherence in Equation~(\ref{evol_imag_coher_lindblad}), and to an analogous decay term for the real component of the coherence in Equation~(\ref{evol_real_coher_lindblad}), with a decay time scale given in both cases by the sum $\Gamma_{ln}$ of the dephasing and relaxation rates.

We emphasize that when energy transfer between the two sites in a dimer is considered, i.e., the smallest possible coupled site system, with only two coupled sites, the non-local coherence terms are not present. 
Then, the reduction of possible coherence effects leads to significantly simpler dynamics, and for a homo-dimer, i.e., $\Delta_{1,2}=0$, Equation~(\ref{evol_real_coher_lindblad}) would even be completely decoupled from the other two equations and the real part of the coherence would simply decay with the decay rate given by $\Gamma_{1,2}$.

\section{Conclusions}
\label{sec_conclusion}

In this paper we have introduced an analysis of electronic excitation energy transport that uses excitation-probability currents to describe the energy transport between the electronic states.
Our discussion was set in terms of a specific focus on systems of coupled sites, such as molecular aggregates, light-harvesting systems, and related systems.
We applied this analysis to an open quantum system model for energy transport, in which the dynamics are described by a non-Markovian quantum master equation that contains both terms representing unitary dynamics and terms that induce dephasing and electronic relaxation, which play an important role in many energy-transfer systems.  
This probability current analysis enabled us to gain instructive insights into key features of the transport.
In particular, it allowed us to identify and quantify the contributions of unitary dynamics, dephasing, and electronic relaxation/dissipation, as well as the contribution of coherence to the transport in a straightforward way.

We found that the probability currents which represent the energy transport are given by a sum of unitary, dephasing, and relaxation contributions.
Each of these have simple forms and can be readily calculated from a given time-dependent electronic density matrix of the system, where this is obtained from, for example, numerical simulation of the quantum master equation. 
An important result of the probability-current analysis is that the unitary contribution to the current is caused entirely by coherence between the sites, i.e., the off-diagonal elements of the electronic system density matrix in the site basis, and not by the population of the sites (diagonal elements of the density matrix).
This means that if there is no coherence between the sites, then the unitary contribution to the current will be zero.
It is also noteworthy that only the {\it imaginary} component of the coherence is involved in the unitary current -- not the real component.
Therefore, it is insightful to consider imaginary and real components of the coherence separately when studying the transport dynamics.

Another important result is that the contribution of the dephasing term to the currents is always zero, regardless of specific properties of the non-Markovian environment, such as, for example, the form of the environment spectral density that describes the strength of the coupling between the electronic excitation and each vibrational mode of the environment.
This means that in a model containing only unitary dynamics and non-Markovian dephasing -- a commonly used model for energy transfer~\cite{cheng2009dynamics, ishizaki2009_17255, ritschel2011absence} -- the only contribution to the overall energy transfer comes from the unitary contribution to the current and thus from the coherence between the sites. 
However, even though the dephasing does not explicitly contribute to the currents, it can affect the coherence that constitutes the unitary contribution to the currents and thereby influence the energy transport implicitly.
If the dephasing is strong, it can destroy the inter-site coherence and thus inhibit the unitary contribution to the transport currents -- this is the well-known Quantum Zeno effect~\cite{rebentrost2009_033003}.  
We note that such a non-Markovian dephasing term in the quantum master equation can also induce relaxation within the vibrational manifold of the environment, since the environment spectral density can be decomposed into single vibrational modes that couple to the system and to a separate environment, leading to relaxation of these single modes~\cite{roden2012accounting}.

We found that the contribution of the electronic relaxation to the current is more complicated and can in general depend explicitly on both coherence and population terms in the electronic density matrix, as well as on the non-Markovian dynamics of the environment.
In the Markov limit, however, when the memory time of the environment is short compared to the other time scales of the dynamics, the electronic relaxation contribution to the current becomes a very simple classical rate equation term that only depends on the populations of the sites and the corresponding relaxation rates, but not on the coherence.  
We note that performing the Markov limit with respect to the electronic relaxation still allows the incorporation of {\it non-Markovian} dephasing -- and hence of {\it vibrational} relaxation as noted above -- caused by a {\it non-Markovian} environment~\cite{roden2012accounting, roden2015psii, kreisbeck2011_2166}.

Such non-unitary electronic relaxation is often used to model radiative or non-radiative decay of electronic excitation or trapping of excitation, e.g.\ in a reaction center of a light-harvesting system~\cite{kreisbeck2011_2166, caruso2009_105106, jesenko2013_174103, roden2015psii}.
This is a separate process from the unitary, coherent excitation transport mechanisms between sites, which occurs, e.g., via resonant transition dipole-dipole interaction between molecules~\cite{may2008charge, ishizaki2009_17255, roden2015psii}.
Thus, even in the presence of electronic relaxation, the actual excitation transport currents between sites often rely on the unitary contribution to the currents, since, as the probability current analysis shows, the third possible contribution, dephasing, does not explicitly contribute to the currents.
In this very commonly encountered situation, the transport currents will then depend entirely on the coherence between the sites.
Consequently, if the coherence is zero at a time $t$, there will be no current, i.e., no energy transport, at this time $t$ and the overall energy transport is dependent on the presence of sustained coherence in the system.

The finding that the energy transport may be entirely determined by the coherence is one of the key insights provided by this probability current approach. 
We emphasize that this reliance of the transport on coherence does not contradict the fact that the (overall) transport dynamics can often be described reasonably well by classical master equations, such as the rate equations deriving from F\"{o}rster and modified Redfield models that contain only population terms and no explicit coherence terms, in contrast to the quantum master equations that contain the coherence as off-diagonal elements of the density matrix~\cite{jesenko2013_174103, raszewski2008_4431, bennett2013_9164, roden2015psii}.
This is because such classical rate equations can be interpreted as {\it effective} descriptions of a full quantum description, in which the effective rate equation {\it implicitly} takes coherence into account by virtue of derived, effective rate parameters, even though the coherence does not explicitly appear in the equation~\cite{jesenko2013_174103}.

Another very useful aspect of the description of energy transport by means of probability currents is that this analysis can reveal the pathways of the transport and their relative importance in a large network of sites and can thus help to gain insight into design-function relationships of biological networks such as light-harvesting complexes.
In this approach, the probability currents between the sites can be integrated over a certain time interval of interest to show the net amount of excitation probability transported via each pathway and the direction of the transport within this time period.
The probability current analysis that we introduced in this paper is used in a companion paper~\cite{roden2015psii} to reveal energy transport pathways in the Photosystem~II super-complex that drives photosynthesis in higher plants, leading to new insights into the design-function relationship of this photosynthetic apparatus. 

We qualitatively analyzed the interplay between the probability current and the imaginary and real components of the coherence for the case of a Markovian environment where the evolution of population and coherence is described by a Lindblad quantum master equation.
This yielded analytic understanding of the mutual influence that these quantities have on each other and how they are affected by dephasing, relaxation, and localization due to energy gaps between the sites -- effects that play a crucial role in modeling and understanding excitation energy transport~\cite{rebentrost2009_033003, chin2010_065002, caruso2009_105106}.

Finally, we point out that while the probability current analysis presented in this paper has focused on a quantum master equation description of energy transfer, our procedure to identify and quantify contributions to the probability currents that stem from the different terms of an evolution equation is more general and can be applied to a broad range of dynamical models and evolution equations. 
It can in principle be applied to any evolution equation that describes the time-dependence of probabilities or other conserved quantities.
This includes the classical master equations that contain probabilities and rates, such as those derived from a generalized F\"{o}rster or modified Redfield models~\cite{raszewski2008_4431, bennett2013_9164}, as well as other forms of quantum master equations.

\begin{acknowledgments}

This material was supported by DARPA under Award No.\ N66001-09-1-2026.

\end{acknowledgments}




\end{document}